\documentclass{aa}
\usepackage{txfonts}
\usepackage{natbib}
\usepackage{ulem} 
\usepackage{amssymb}    
\usepackage{amsfonts}   
\usepackage{bm}

\def\dd{\mathrm{d}}

\bibpunct{(}{)}{;}{a}{}{,} 

\begin{document}

\title{Can dark energy emerge from quantum effects in a compact extra
  dimension?}
\titlerunning{Dark energy as quantum vacuum?}

\subtitle{} 

\author{Arnaud Dupays\inst{1,3,4}
  \and Brahim Lamine \inst{2,3,4}
     \and Alain Blanchard \inst{3,4}} 

\authorrunning{A. Dupays et al.}

\offprints{B. Lamine, \email{brahim.lamine@irap.omp.eu}}

\institute{Universit\'e de Toulouse, UPS, Laboratoire Collisions
  Agr\'egats R\'eactivit\'e, IRSAMC, F-31062 Toulouse, France \and
  Universit\'e Pierre et Marie Curie, UPMC, ENS,
  Laboratoire Kastler Brossel, F-75252 Paris Cedex 05, France \\
  CNRS, UMR 8552, F-75252 Paris, France
  \and Universit\'e de Toulouse, UPS-OMP, IRAP, Toulouse, France\\
  \and CNRS, IRAP, 14, avenue Edouard Belin,, F-31400 Toulouse, France}

\date{ \today }

\abstract{ The origin of the accelerated expansion of the universe is
  a major problem in both modern cosmology and theoretical physics. In
  quantum field theory, simple estimations of the vacuum contribution
  to the density energy of the universe are known to lead to
  catastrophically high values compared to observations. A
  gravitational Casimir effect from an additional compact dimension of
  space is known to lead to an effective cosmological
  constant. Nevertheless, such a contribution by itself is usually not
  regarded as a plausible source for accelerating the expansion, given
  the constraints on such scenarios. There, we propose that the
  Casimir vacuum contribution of the gravitational field actually
  provides a low positive value to the density energy of the
  universe. The key new ingredient is to assume that only modes with
  shorter wavelengths than the Hubble radius contribute to the vacuum
  energy. Such a contribution gives a positive energy density, has a
  naturally Lorentz invariant equation of state in the usual 4D
  spacetime, and can thus be interpreted as a cosmological
  constant. Its value agrees with observations for a radius of a fifth
  extra dimension given by $35\,\mu$m. The implied modification of the
  gravitational inverse square law is close but below existing limits
  from experiments testing gravity at short range.  }

\maketitle

\section{Introduction}

The evidence of the accelerated expansion of the universe has grown
since the first result from the Hubble diagram of distant type Ia
supernovae~\citep{Riess1998,Perlmutter1999}. The angular power
spectrum of the fluctuations in the cosmic microwave background and
the large-scale properties of the galaxy distribution are all
consistent with the accelerated expansion of a homogenous universe,
while no alternative Friedmann-Lema\^{\i}tre model seems to be able to
reproduce these three data
sets~\citep{Frieman2008,Blanchard2010}. Dark energy, the origin of the
cosmic acceleration, is often qualified as one of the deepest
mysteries in modern physics, and its origin is hard to explain within the
standard framework of high-energy physics~\citep{Weinberg:1989}. This
issue has been a tremendous stimulation for the community, producing a rich
ensemble of theoretical approaches, while being the target of
unprecedented efforts in astrophysical observational strategy, either in
the form of ground projects~\citep{LSST} or ambitious space projects
like EUCLID~\citep{Euclid}.

A genuine cosmological constant $\Lambda$, as introduced by Einstein
in 1917, accounts for the observed cosmic acceleration. However, most
scientists agree on the lack of theoretical motivation for introducing
such a term into the Einstein equations~\citep{Amendola:2012}. One
reason that is often invoked is that $\Lambda$ introduces a new
fundamental energy scale if one introduces the Planck constant
$\hbar$:
\begin{equation}
  E_\Lambda=\left((\hbar c)^3\rho_\Lambda\right)^{1/4}\simeq2\;\text{meV}\,,
\end{equation}
with $\rho_\Lambda=\Lambda c^4/(8\pi G)$ representing 73\,$\%$ of the
energy content of the Universe. On this energy scale, no exotic
physics are a priori expected. Equivalently, a dimensional length
scale can be associated:
\begin{equation}
\ell_\Lambda =\left(\frac{\hbar c}{\rho_\Lambda}\right)^{1/4}
\simeq 83\;\mu\text{m}\,.
\end{equation}
An experimental effort has been devoted to observing any deviation
from the gravitational laws~\citep{Adelberger:2009} on this scale and
below. No anomaly has been observed just below this length
scale~\citep{Kapner:2007}. Of course, as we are going to see, this
does not exclude any deviation in the gravitational laws controled by
$\ell_\Lambda$, because numerical factors could lower the true length
scale to a value below $\ell_\Lambda$. Finally, we need to mention
that if $\Lambda$ is a true fundamental constant, then it is not
possible to define a single natural length scale from $\hbar$, $c$,
$G$, and $\Lambda$, but instead one can have
\begin{equation}
  \label{eq:11}
  \ell=\Lambda^{-1/2}f\left(\frac{\hbar G\Lambda}{c^3}\right)
\end{equation}
with $f$ an arbitrary function of the dimensionless constant
$\frac{\hbar G\Lambda}{c^3}\sim2\times10^{-122}$. This low a number is
nothing more than a reformulation of the cosmological constant
problem. Taking $f(x)=1$ leads to a cosmological scale for $\ell$ (the
size of a static Einstein universe), and $f(x)=\sqrt{x}$ leads to the
Planck length, while $f(x)=x^{1/4}$ gives the previously introduced
scale $\ell_\Lambda$, qualified as the natural dark-energy length
scale. This scale is the geometric mean of the two former
scales.

Historically, a physical explanation for the cosmological constant
came from identification of this term with a Lorentz invariant
vacuum~\citep{Lemaitre:1934}, which leads to the possibility of a
gravitationally active vacuum due to the contribution of zero-point
energy. This attractive idea was discussed as early as in the 1920s by
Nernst and Pauli
(see~\citet{Straumann:2002,PeeblesRatra2003,Kragh:2011} for a
historical presentation) but it was immediately realized that this
possibility is plagued by a large discrepancy in the estimated order
of magnitude. To avoid dramatic consequences for cosmology, it is
usually assumed that those vacuum energies do not gravitate or give a
renormalized value that is exactely zero. This is the first
cosmological constant problem. We are therefore left with the second
cosmological problem of how to explain the low ``incremental''
positive value observed today. A first original idea has been
historically proposed in~\citet{ZelDovich:1967,Zeldovich:1968},
considering $\rho_\Lambda$ as being gravitational interaction energy
between virtual pairs of the QED vacuum. Unfortunately, this elegant
propositon could still not explain the low value of the possible
cosmological constant.  Nowadays, the actual contribution of vacuum to
the present-day density of the universe is still the subject of debate
in the scientific community.

Here, we focus on the possibility of identifying the cosmological
constant with effects from the quantum vacuum. In section 2, we
briefly discuss the vacuum contribution problem. In four-dimensional
spacetime, it can be argued that this contribution vanishes for a
massless field, while for a massive field the expected contribution is
much greater than the observed one for any standard fundamental mass
scale unless fine tuning is invoked. In section 3. we examine the case
of adding extra spatial compact dimensions. Indeed, pioneer papers in
the 80s~\citep{Appelquist:1983a,Appelquist:1983b} computed the quantum
corrections in the energy density of the vacuum stemming from the
presence of such extra dimensions. It has been shown that those
quantum corrections correspond to a Casimir effect of the
gravitational field induced by the periodic boundary conditions along
the extra compact dimensions~\citep{Rohrlich:1984}, but direct
identification of this quantum correction with the cosmological
constant~\citep{Milton:book,Gardner:2002,Elizalde:2006a} encountered
some difficulties that we recall briefly later. However, we show that
including the Hubble scale as a maximum wavelength allowed for quantum
vacuum modes of the gravitational field provides a mechanism to
generate a positive cosmological constant with one extra dimension. We
furthermore show that such a scenario leads to a modification of the
gravitational inverse square law on scales that should be accessible
by experiments in the near future.

\section{The zero-point energy contribution to the vacuum }

Considering the example of a massive scalar field, the contribution of
zero-point energy to the density can be obtained as the vacuum
expectation value of the 00 component of the energy momentum tensor
$T^{\mu\nu}$ ($\hbar=c=1$)
\begin{equation}
  \rho_v =  \langle 0\vert T^{00}\vert0\rangle = 
\int\frac{\dd^d\bm{k}}{(2\pi)^d}\frac{1}{2}\sqrt{\bm{k}^2+m^2}
\label{eq:rhov}
\end{equation}
with $d$ the number of spatial dimensions and $k$ the wave vector. The
vacuum pressure can be computed in a similar way to the density:
\begin{equation}
\label{eq:pv}
p_v = (1/d)\sum_i \langle 0\vert T^{ii} \vert 0\rangle = 
\frac{1}{d}\int\frac{\dd^d\bm{k}}{(2\pi)^d}\frac{1}{2}\frac{\bm{k}^2}
{\sqrt{\bm{k}^2+m^2}}\,.
\end{equation} 

These contributions are highly divergent and therefore need some
regularization treatment. The most trivial regularization procedure
would be to introduce an ultraviolet cutoff $k_c$ in momentum space,
above which the theory breaks down. Nevertheless, this procedure
introduces two flaws: i) the energy density scales as $k_c^{d+1}$,
which leads to a catastrophic value compared to the observed energy
density in our universe for any scale $k_c$ related to high-energy
physics scales, ii) this cutoff in momentum explicitly violates
Lorentz-invariance and leads to a vacuum expectation value of the
energy-momentum tensor, which is not proportional to $g_{\mu\nu}$ and
therefore cannot be accepted as such for a description of vacuum. The
inclusion of non-Lorentz invariant counter terms can restore the
symmetry and lead to the correct equation of
state~\citep{Hollenstein:2011}. Another convenient approach is to use
a covariant regularization, such as the dimensional regularization in
which the number of dimensions $d$ is written as $d=D+\epsilon$, with
$D$ an integer and $\epsilon\rightarrow0$. Introducing a constant
$\mu$ (the dimension of which is a mass, or the inverse of a
length) so that the energy density and pressure keep the correct
dimension, one obtains (see for example~\citet{Martin:2012})
\begin{equation}
  \label{eq:1}
  p=-\rho=\frac{m^{d+1}\Gamma\left(-\frac{d+1}{2}\right)}
{\mu^{\epsilon}2^{d+2}\pi^{\frac{d+1}{2}}}\,.
\end{equation}
For instance, for $D=3$, discarding the diverging $1/\epsilon$ term
and using the modified minimal subtracting scheme, one finally obtains
\begin{equation}
  \label{eq:5}
  p=-\rho=-\frac{m^4}{64\pi^2}\,\ln\left(\frac{m^2}{\mu^2}\right)\,.
\end{equation}
It is now explicit that the Lorentz-invariance is preserved (since
$p=-\rho$). Moreover, the scaling of the energy density is now like
$m^{d+1}$, which is better than $k_c^{d+1}$ in the hard cutoff
regularization. Nevertheless, the presence of the regulator $\mu$ does
not allow a prediction for $\rho$, while natural values for $\mu$ lead
to catastrophically high value compared to the observed value of
$\rho_\Lambda$. In any event, the important point to be stressed at
this level is that for a massless field ($m=0$) the contribution to
the vacuum energy density is exactly zero so that this regularization
procedure accounts for a degravitation of massless fields, even if it
does not give any physical mechanism that would be at its origin
(see~\citet{Ellis:2011,Smolin:2009} for one example of such
theories). This result corroborates the simple remark that if we were
to build a traceless energy momentum tensor from the metric
$g^{\mu\nu}$, the only solution would be to have $\langle
T^{\mu\nu}\rangle=0$. Said differently, to accomodate the equation of
radiation (i.e. massless fields), $p=\rho/d$ with the one of vacuum,
$p=-\rho$, one needs $p=\rho=0$.

Though the specific consideration for a massless field does not stand
for a general demonstration, the previous consideration corroborates
the standard conclusion that some mechanism sets the contribution of
vacuum energy to exactly zero in an isotropic spacetime of arbitrary
dimensions. The origin of the accelerated expansion of the universe is
then logically expected to happen by means of a distinct physical
mechanism. The late domination of a scalar field or modifications to
the Einstein-Hilbert action are the two options most investigated so
far, and have been the subject of intensive research activities ever
since there is evidence of an accelerated
expansion~\citep{Clifton:2012,Tsujikawa:2010}.

In the next section we present a physical mechanism to generate a
nonzero positive density energy and pressure from zero-point energies
of a massless field (the gravitational field itself). This is achieved
by assuming the existence of an additional compact spatial dimension,
which will therefore modify equation (\ref{eq:1}). It is well known
that modification of the boundary conditions of a quantum field leads
to nontrivial physical properties of the vacuum. The Casimir force
between two infinite conducting plates is a famous example of a
physical nonzero but finite contribution from the QED vacuum even if
the electromagnetic field is massless. In the latter configuration,
the isotropy of space has obviously been broken by the presence of
boundary conditions. The pressure in the direction normal to the
plates satisfies $p_\perp=3\rho$ (with $\rho < 0$), while the pressure
parallel to the plates satisfies
$p_\parallel=-\rho$~\citep{Brown:1969}, in accordance with the
traceless nature of the electromagnetic field. Remarkably enough, the
Lorentz invariance in the two dimensions parallel to the plates ensures
the equation of state $p_\parallel=-\rho$ with a nonzero value of
$\rho$. As we will see, in the presence of additional compact
dimensions of space, a gravitational Casimir effect allows for a
nonzero density energy that is Lorentz invariant in the usual 4D
spacetime ($p=-\rho$), even for a massless (traceless) field.

\section{Casimir effect from a higher compact dimension}

The existence of additional space dimensions has been considered with
various purposes in modern physics (for a review, see for
example~\citet{Rubakov:2001}), from the Kaluza-Klein
scenario~\citep{Kaluza:1983} aiming at unifying interactions to the
more recent braneworld paradigm dealing with the hierarchy
issue~\citep{Arkani:1998,Antoniadis:1998,Randall:1999}. In this
picture, matter is localized in a 4D spacetime (the brane), while
gravity can propagate in all the dimensions (the bulk). This picture
allows for a large extra dimension (see also~\citet{Antoniadis:1990}
for a first proposal using large extra dimensions). Because the
gravitational field is massless, dimensional regularization
(equation~(\ref{eq:1})) ensures that the energy density vanishes in
arbitrary $N$-dimensional infinite isotropic spacetime.  However, in
the case of compact additional dimensions, the situation is different
since the structure of the quantum vacuum is modified by the
quantification of the gravitational field in the additional
dimensions. This quantification of the gravitational-field modes in
the bulk leads to a Casimir energy that was computed many years
ago for one extra dimension in pioneer works from the 80s
in~\citet{Appelquist:1983a,Appelquist:1983b} for a Minkowski
background metric and later in~\citet{Rohrlich:1984} using a
zero-point energy calculation and an exponential cutoff
regularization.

In what follows, we first reproduce the calculation for $N=1$ using
dimensional regularization. We thus assume the existence of one
spatial additional dimension compactified on a circle of radii
$R$. The periodic condition $f(x^i,x^4+2\pi R)=f(x^i,x^4)$ allows the
metric tensor to be expanded in Fourier series:
\begin{equation}
  \label{eq:2}
  g_{\mu\nu}(x^i,x^4)=\sum_{n=-\infty}^{\infty}g^{(n)}_{\mu\nu}(x^i)
\exp\left(inx^4/R\right)
\end{equation}
where $x^4$ is the coordinate in the extra dimension. In a specific
gauge, the metric satisfies the propagation equation
$\nabla^2g_{\mu\nu}=0$ so that the gravitational modes
$g^{(n)}_{\mu\nu}$ satisfy the dispersion relation:
\begin{equation}
  \label{eq:disp}
 \omega_n({\bm{k}})=\sqrt{{\bm{k}}^2+n^2/R^2}\,.
\end{equation}
The mode $n=0$ is the usual massless graviton, while the excited modes
$n\neq0$ correspond to effective massive gravitational fields of
masses $n/R$ (Kaluza-Klein tower). To simplify, we model the
gravitational field by a scalar field and multiply the final result by
the number of polarization states $p_m=m(m-3)/2$ in $m$-dimensional
spacetime ($p_5=5$). The previous assumption is justified since we
consider a flat extra dimension. For situations with curvature, a
conformally coupled scalar field would have been a better description
of the true gravitational field.

With a vacuum energy per mode given by $\omega_n/2$, the total vacuum
energy density is obtained as
\begin{equation}
  \label{eq:qvh}
  \rho=\frac{p_{d+2}}{2\pi R}\sum_{n=-\infty}^{\infty}\;\int\frac{\dd^d\bm{k}}{(2\pi)^3}
\frac{1}{2}\sqrt{{\bm{k}}^2+n^2/R^2}\,.
\end{equation}
Each term in the previous sum can be dimensionally regularized using
eq. (\ref{eq:1})~:
\begin{equation}
  \rho=-\frac{2 p_{d+2}\Gamma\left(-\frac{d+1}{2}\right)}
{(4\pi)^{\frac{d+3}{2}}R^{d+2}}\,\zeta_R(-d-1)
\end{equation}
where $\zeta_R$ is the Zeta Riemann function. Using the reflection
formula
\begin{equation}
  \label{eq:10}
  \Gamma\left(\frac{z}{2}\right)\zeta_R(z)\pi^{-z/2}=\Gamma\left(
\frac{1-z}{2}\right)\zeta_R(1-z)\pi^{(z-1)/2}\,,
\end{equation}
one finally obtains a finite (regularized) contribution for $d=3$
\begin{equation}
\rho_{\text{App}}=-\frac{15\zeta_R(5)}{128\pi^7R^5}\simeq-4\times10^{-5}
\frac{1}{R^5}\,.
\label{eq:app}
\end{equation}
This expression agrees with previous studies based on different
regularization schemes: hard cutoff in~\citet{Appelquist:1983a},
exponential cutoff in~\citet{Rohrlich:1984}, or point-splitting
in~\citet{Milton:book}. The density in the brane is then obtained by a
trivial integration (in the fifth dimension):
 \begin{equation}
   \rho_{\text{brane}}=-\frac{15\zeta_R(5)}{64\pi^6R^4}\simeq-2.5\times10^{-4}
   \frac{1}{R^4}\,.
\label{eq:app2}
\end{equation}

Generalizations to a spacetime structure $M^4\times S^N$ have been done
in~\citet{Candelas:1984} and~\citet{Chodos:1985} for odd $N$ and later
in~\citet{Myers:1986} and~\citet{Kantowski:1987} for even $N$. Such
generalizations to extra dimension $N$ can be written as
 \begin{equation}
\rho_{\text{brane}}=\kappa_N\frac{\hbar c}{R^4}
\label{eq:appN}
\end{equation}
and are summarized in Table~\ref{table:1}. The second colum gives the
relation between the gravitational Casimir energy $\rho$ and the
radius of the extra dimension in $R^4\times S^N$ compactification for
different values of $N$ (the differences with~\citet{Candelas:1984}
and ~\citet{Chodos:1985} being the polarization factors
$p_{d+2}$). The third column gives the radius of the extradimension,
such that this gravitational Casimir energy is equal to the observed
dark energy density (the sign being the one of the normalization
constant in column 2). The fourth column gives the size of the
additional dimension that would solve the hierarchy problem (i.e. in
order to have a Planck scale equal to $1$~TeV). Finally, the last
column summarizes the present observational constraints on the size of
this extra dimension, from $N=1$ to $N=7$ extra
dimensions~\citep{Beringer:2012,Fermi:2012}. For instance, one can see
from this table that the hierarchy problem can only be solved with $N
\geq 6$. The conclusion to be drawn from this table is that extra
dimensions cannot solve the hierarchy problem and explain the origin
of dark energy at the same time without any additional ingredient. We
now assume that the extra dimension is not introduced to solve the
hierarchy problem. Then, if one identifies the cosmological constant
with the Casimir energy, it is clear that odd values of $N$ are still
excluded. Even values of $N$ greater than one are more problematic,
since the evaluation of their contribution contains a logarithmic term
of some unknown scale $\mu$. This makes the normalization constant
(column 2) not well determined. Nevertheless, any plausible value of
$\mu$ (say below the Planck mass) will not make a large numerical
difference and will therefore lead to a radius that is not very different
from those obtained in the odd case. Therefore we are left with the
only possibility of having one extra dimension if the cosmological
constant has to be created only from the Casimir energy contribution
of a higher compact dimension. However, in the 1D case, as can be seen
from Table~\ref{table:1}, a negative sign then seems to be obtained
for $\rho$, while observations request a positive sign. More
sophisticated and somewhat speculative scenarios have been proposed to
overcome this dead end, although no completely convincing solution has
emerged~\citep{Milton:2003,Elizalde:2006b}. Among those scenarios, we
mention the possibility of considering a nontrivial topology of
space~\citep{Elizalde:2006a,Elizalde:2006b} or different boundary
conditions (twisted boundary condition will change the sign of the
Casimir energy~\citep{Milton:2003}, as well as $\mathbb{Z}^2$ symmetry
on a torus compactification $T^2$ with certain shape moduli $\theta$,
as shown in~\citet{Matsuda:2006}). They also comprise new fermion
fields in the bulk~\citep{Greene2007,Matsumoto2013}, massive
graviton~\citep{Bauer:2005}, a latticed
extradimension~\citep{Cognola2004,Cognola:2005}, a time evolution of
the size of the extra dimension~\citep{Ponce:2005,Blanchard2012} or
even of the scale factor~\citep{Cahill:2011,Bernard2013}, curvature in
the bulk (for example anti-de Sitter and a massive scalar field as
in~\citet{Li:2005}, or a bulk fermion field in the Randall-Sundrum
model~\citep{Shao:2010}). In conclusion, previous attemps to directly
identify this Casimir energy with the cosmological constant have not
led to a definite conclusion.

\begin{table*}
  \centering
\noindent\begin{tabular}{|c|c||c|c|c|}
\hline $N$ &$\kappa_N=\frac{R^4\rho}{\hbar c}$&
\begin{tabular}{c}$R^\Lambda_N$\\$(\mu \text{m})$\end{tabular}
&\begin{tabular}{c}$R^{\text{hierar.}}_N$\\$(\mu \text{m})$\end{tabular}
&\begin{tabular}{c}Constraints\\($\mu$m)
\end{tabular}\\
\hline 1&$-2.5\times10^{-4}$&$(10.5)$&$2.6\times10^{19}$&
\begin{tabular}{c}$< 44 \text{ (ISL)}$\\$< 44 \text{ (NS)}$\end{tabular}\\
2&--&--&$2.2\times10^{3}$&
\begin{tabular}{c}$< 30 \text{ (ISL)}$\\$< 0.00016 \text{ (NS)}$ 
\end{tabular}\\
3&$1.1\times10^{-3}$&$15.0$& $9.7\times10^{-3}$&
\begin{tabular}{c}$< 2.6\times10^{-6}
  \text{ (NS)}$\\
  $< 10^{-3}\text{ (LHC)}$\end{tabular}\\
4&--&--&$2.0\times10^{-5}$&
$< 3.4\times10^{-7} \text{ (NS)}$\\
5&$1.2\times10^{-2}$&$27.2$& $5.0\times10^{-7} $&$< 1.0\times10^{-7}
\text{ (NS)}$\\
6&--&--&$4.3\times10^{-8}$&
$< 4.4\times10^{-8} \text{ (NS)}$\\
7&$3.6\times10^{-2}$&$36.1$&  $7.3\times10^{-9}$&$< 2.4\times10^{-8} \text{
  (NS)}$\\
\hline
\end{tabular}
\caption{Summary of constraints on extra dimensions. ISL is for 
  inverse square law
  test~\citep{Adelberger:2009}, NS for neutron star
  contraints~\citep{Hannestad:2003}, and LHC from the CMS experiment at
  CERN.}
\label{table:1}
\end{table*}

In the following we re-examine this question in the cosmological
context. We show that this provides a mechanism leading to a positive
Casimir energy density $\rho$ at late times. The key ingredient is to
take the finite age of the Universe into account. This finite age
implies the existence of a length scale, the Hubble radius
$cH(t)^{-1}$. This clearly adds a boundary condition that has to be
considered. We make two assumptions to account for this effect. First,
only modes corresponding to wavelengths shorter than the Hubble radius
$cH(t)^{-1}\sim ct$ contributes to the density of the vacuum energy
(see also~\citet{Cahill:2011} for a similar proposition in the
cosmological context, or~\citet{padmanabhan:2012} for a different
mechanism also implying the Hubble radius). The second assumption is
that as long as the Hubble radius is shorter than the radius $R$ of
the extra dimension, the energy density is equal to zero. The reason
is that when the horizon is smaller than the radius of the extra
dimension, the structure of the quantum vacuum cannot depend on the
compact nature of the extra dimension because gravitons have not yet
explored the ``compactness'' of space. The situation should therefore
be equivalent to the one previously discussed of a massless scalar
field in an isotropic spacetime, leading to $\langle
T_{\mu\nu}\rangle=0$ (see Eq.~(\ref{eq:1})). It is easy to see why
those assumptions can yield a net positive contribution of zero-point
energy.  Indeed, when the horizon radius crosses the radius of the extra
dimension, the change in the vacuum is only due to new modes that
appear with a wavelength larger than $2\pi R$. Those modes contribute
with $\hbar\omega/2$ of vacuum energy and a UV cutoff of about $1/R$,
leading to a finite positive contribution. In this picture, the
cosmological constant can be seen as a ``temporal'' Casimir effect, as
if the boundary conditions were switched on at a given moment of
time. The observable quantity is therefore the change of vacuum
energy when the Hubble radius crosses the extra dimension.

The previous discussion implies that (\ref{eq:app}) has to be changed
in order to fix the subtration point in the energy density at
$t=R/c$. To perform this task, we add a low-energy cutoff
$\omega_n(\bm{k})>2\pi/t$ to (\ref{eq:qvh}) and a counterterm $CT(t)$,
which restores Lorentz invariance and insures that $\rho$ is zero as
long as $t\leq R/c$,
\begin{equation}
  \rho(t)=\frac{5}{R}\int_{\omega_n(\bm{k})>2\pi/t}\frac{\dd^3\bm{k}}{(2\pi)^3}
  \sum_{n=-\infty}^{\infty}\;
  \frac{1}{2}\sqrt{{\bm{k}}^2+n^2/R^2} +CT(t)
\end{equation}
with $CT(t)$ such that $\rho(t\leq 2\pi R/c)=0$. At later times, the
boundary condition changes and the energy density is no longer kept
to zero. The counterterm then stays equal to its value at time $2\pi
R/c$ (which we note $CT$), obtained from the transition condition
$\rho(2\pi R/c) = 0$
\begin{equation}
\label{eq:horizoncondition}
\frac{5}{R}\int_{\omega_n(\bm{k})>0}(\ldots)-\frac{5}{R}
\int_{\omega_n(\bm{k})<c/R}(\ldots) +CT= 0\,.
\end{equation}
In the late time regime, $t\gg R/c$, the cut off introduced by the
Hubble radius can be neglected so that the present-day energy density
$\rho_0$ reads as
\begin{eqnarray}
  \label{eq:qvh0}
\nonumber  \rho_0 & = & \frac{5}{ R}\int_{\omega_n(\bm{k})>0}
\frac{\dd^3\bm{k}}{(2\pi)^3}\sum_{n=-\infty}^{\infty}\;
\frac{1}{2}\sqrt{{\bm{k}}^2+n^2/R^2} +CT\\
& = &
\frac{5}{ R}\int_{\omega_n(\bm{k}) < c/R}\frac{\dd^3\bm{k}}{(2\pi)^3}\sum_{n=-\infty}^{\infty}\;
\frac{1}{2}\sqrt{{\bm{k}}^2+n^2/R^2} 
\end{eqnarray}
after using equation (\ref{eq:horizoncondition}). We see in the
previous expression that $R$ acts as a UV cutoff for the sum of
zero-point energies. The condition $\omega_n(k)^2 < (1/R)^2$ implies
that only the term $n=0$ contributes to the last integral, rendering it
elementary. This allows us to obtain the value of the density
(reintroducing explicitely $\hbar$ and $c$):
\begin{equation}
  \label{eq:cc}
  \rho_0=\frac{5\hbar c}{32\pi^3 R^5}\,.
\end{equation}
The other components of the energy-momentum tensor can be obtained
from $\rho_0$. Indeed, the traceless nature of the gravitational field,
together with the symmetry of the problem, requires that
\begin{equation}
  \label{eq:12}
  \langle T^{\mu\nu}\rangle=\rho_{\text{cas}}(g^{\mu\nu}+5\hat{n}^\mu
\hat{n}^\nu)
\end{equation}
with $\hat{n}^\mu$ the unit spacelike vector pointing in the extra
dimension ($\hat{n}^2=-1$) and $\rho_{\text{cas}}$ is a constant
(because of the conservation laws $\partial_\mu T^{\mu\nu}=0$). One
finds that the pressure in the extra dimension (perpendicular to the
brane) is $p^\perp=4\rho_0$, while the pressure in the brane (the
usual spacetime dimension) is such that $p^\parallel=-\rho_0$. This
situation is analogous to the previous discussion of the
electromagnetic Casimir situation (section 2.). Also $p^\perp$ could
have been derived from energy conservation when considering a
variation in the radius $R$. On the brane, the energy-momentum tensor
is obtained by integrating over the fifth dimension,
\begin{equation}
  \label{eq:4}
  \rho_{\text{brane}} = \frac{5\hbar
  c}{16\pi^2 R^4}\quad,\quad p_{\text{brane}}=-\rho_{\text{brane}}
\end{equation}
Equation (\ref{eq:4}) can thus be identified with the present-day dark
energy density $\rho_{\text{DE}}=0.7\rho_c\approx 4$ keV/cm$^3$ for an
appropriate value of $R$. Such an identification leads to predicting
the size of the extra dimension given by
\begin{equation}
  \label{eq:9}
  R=\left(\frac{5\hbar G}{2\pi c^3\Lambda}\right)^{\frac{1}{4}}=
35\;\mu\text{m}\,.
\end{equation}
A consequence of the present discussion is that gravitational laws are
modified on the scale of the radius $R$, which is precisely the range
of present experiments, such as the inverse square law
tests~\citep{Kapner:2007,Adelberger:2009} or experiments aiming at
measuring the Casimir force~\citep{Antoniadis:2011}. More
interestingly, because of numerical prefactors, the value of $R$
predicted here is slightly lower than the dimensional length scale
$\ell_\Lambda$ introduced in the introduction.

The 4D gravitational potential, in the presence of one
extra dimension is obtained as an infinite sum of Yukawa potentials,
each of them corresponding the one massive mode of the Kaluza Klein
tower~\citep{Arkani:1999,Kehagias:2000}
\begin{equation}
  \label{eq:6}
  V=-\frac{G_3M}{r}\sum_{m=-\infty}^\infty
  e^{-\vert m\vert\frac{r}{R}}=
  -\frac{G_3M}{r}\,\coth\left(\frac{r}{2R}\right)\,.
\end{equation}

\begin{figure}
  \centering
  \includegraphics[width=1.0\linewidth]{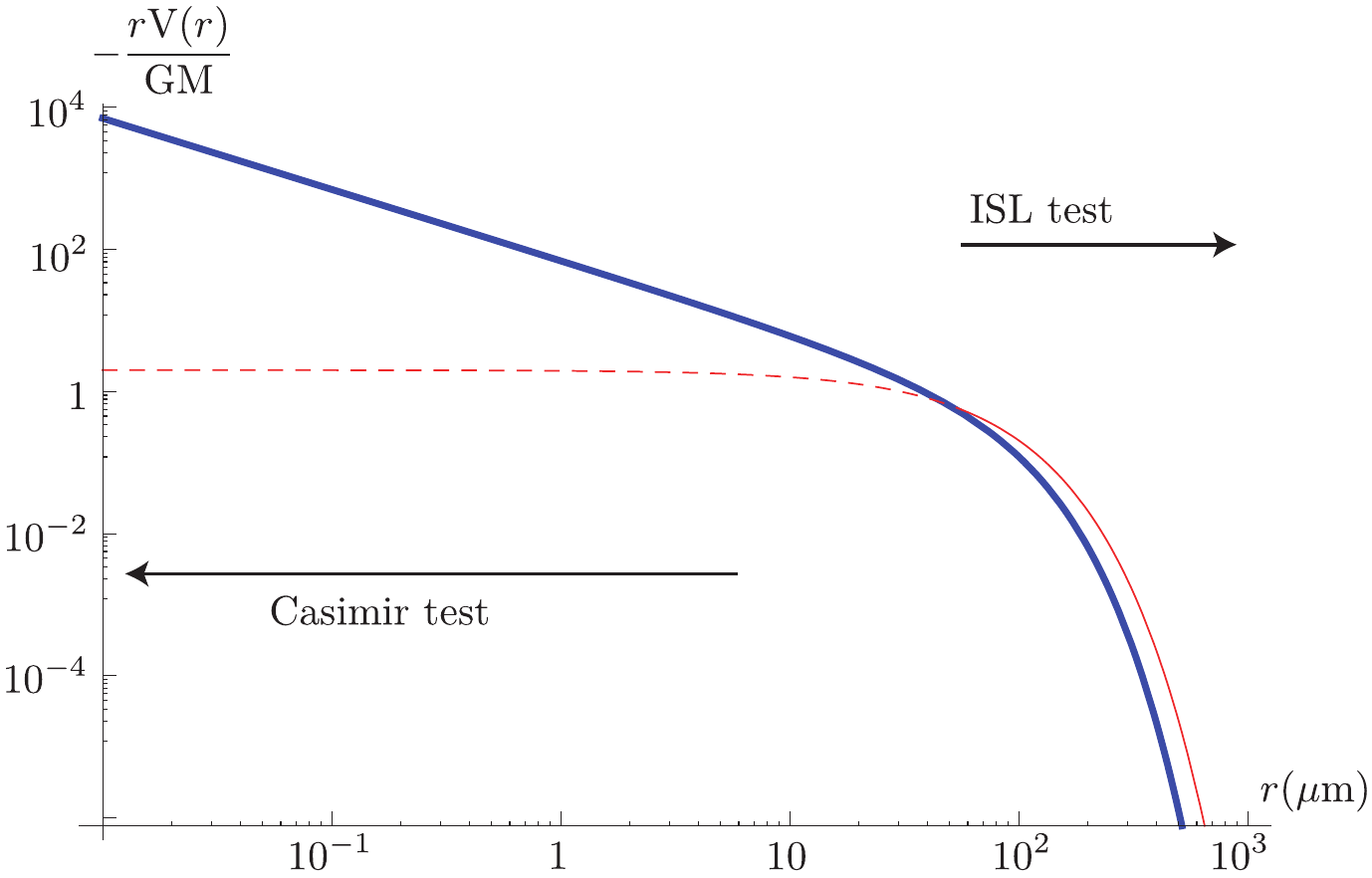}
  \caption{Point-particle gravitational potential for one extra
    dimension (bold black line) and Yukawa (red line). The Yukawa
    modification is taken with a range given by the present-day
    constraint stemming from~\citet{Adelberger:2009},
    $\lambda=44\,\mu$m. The dashed part of this curve corresponds to
    scales not tested in~\citet{Adelberger:2009}. The extradimension
    potential is plotted for a value of the radius given by
    $R=35\,\mu$m. Our prediction is not excluded by experiments, but
    further improvement will soon give a definite answer. The
    short-scale behavior is different from the pure Yukawa
    modification usually searched for in experiments.}\label{fig:1}
\end{figure}

For $r\gg R$, the previous expression is given by the Newtonian
expression ($m=0$) plus the contribution of the lightest Kaluza-Klein
modes ($n=\pm1$),
\begin{equation}
  \label{eq:7}
V\simeq-\frac{G_3M}{r}(1+2\exp(-r/R))  \qquad,\qquad r\gg R\,.
\end{equation}
This corresponds to a Yukawa modification with strength $\alpha=2$ and
a range given by the radius $R$ of the extra dimension. Using this
type of potential, the analyses of ISL tests~\citep{Adelberger:2009}
give a maximum size of $44\;\mu$m for $R$ at $95\%$ confidence
level. Our prediction is therefore just below the present-day
limits. Improvement of these measurements will therefore be critical
for testing our model; nevertheless, when probing the ISL at distance
$\sim R$, the complete expression should be used instead of the simple
Yukawa description (see figure \ref{fig:1}). On smaller scales, the
best constraints on gravity laws are obtained by Casimir force
measurements~\citep{Decca:2007}. The experiments are performed at a
smaller distance than the size of the extra dimension, leading to a
different behavior for the potential (\ref{eq:6})
\begin{equation}
  \label{eq:8}
  V\simeq -\frac{2G_3MR}{r^2}\,.
\end{equation}

It leads to a power-law modification of the gravitational force
between two test masses with an amplitude scaled by $R$ given by
eq. (\ref{eq:9}). This modification could be searched for in Casimir
experiments operating at short distances, although present-day limits
in those experiments are still several orders of magnitude above our
prediction~\citep{Antoniadis:2011}.

\section{Conclusion}

The zero-point energy from a quantized field present in an additional
compact dimension naturally provides a nonvanishing value for the
vacuum contribution to density of the universe, through a Casimir-like
effect. Such a term is naturally Lorentz invariant in the usual 4D
spacetime and therefore may provide a natural explanation for the
observed cosmological constant. However, present-day experimental
limits on possible additional dimension, summarized in Table 1,
exclude more than one extra dimension for such contribution to be the
only origin of the observed dark energy density~\footnote{Strictly
  speaking this conclusion holds only for odd number of extra
  dimensions, as the actual Casimir contribution for even number of
  extra dimensions is not properly known. Of course, if other
  contributions are present (brane tensions, warping etc...), large
  extra dimensions are still viable.}. The case of one extra dimension
is still allowed, although in the case of a Minkowski spacetime, it
leads to a negative contribution to the density of the universe. 

In the cosmological context, we have proposed that the Hubble radius
acts as a cut-off to mode wavelenths contributing to the vacuum
expectation value, i.e. an infrared cut-off in eq.
(\ref{eq:rhov}). Using a zero-point energy calculation, we showed that
this infrared cut-off yields a positive contribution. Therefore this
mechanism naturally explains the origin of the observed cosmic
acceleration that appears as a manifestation of the quantized
gravitational field in an additional dimension. A first consequence of
this model is that the Planck energy scale is lowered to $\sim10^9$
GeV. A second consequence is that the equation of state of
cosmological dark energy should be exactly that of a cosmological
constant, i.e. $w = -1$. A third consequence is that gravitation law
would be modified on scales that are about the size of the compact
dimension, which is $35\,\mu$m, a value below the purely dimensional
dark energy length scale $(\hbar
c/\rho_v)^{1/4}\sim85\,\mu$m~\citep{Kapner:2007, Beane:1997} and below
but close to present experimental limits on departure to the inverse
square law of gravity at short scales. This leaves open the
fascinating possibility that tests of the gravitation law on short
distance shed new light on the nature and origin of cosmic
acceleration.

\begin{acknowledgements}
  The authors are grateful to Serge Reynaud, Carlo Rizzo, and Bertrand
  Chauvineau for fruitful discussions.
\end{acknowledgements}

\bibliographystyle{aa} 
\bibliography{biblio_AA} 

\end{document}